\documentclass[aps,prr,twocolumn,superscriptaddress,amsmath]{revtex4-2}	
\usepackage{graphicx}       
\usepackage{color}
\usepackage[colorlinks,urlcolor=blue,citecolor=blue,linkcolor=blue]{hyperref}
\usepackage{cancel}
\usepackage{orcidlink}
  
\graphicspath{{figures_manuscript/}}
\usepackage{verbatim}  
\usepackage[normalem]{ulem}     
 
\begin{document}  

\title{Time-dependent spectra of quantum beats}

\author{H\'ector M. Castro-Beltr\'an\,\orcidlink{0000-0002-3400-7652}} 
\email{hcastro@uaem.mx}
\affiliation{Centro de Investigaci\'on en Ingenier\'ia y Ciencias
Aplicadas, Universidad Aut\'onoma del Estado de Morelos, 
Avenida Universidad 1001, 62209 Cuernavaca, Morelos, Mexico}
\author{Ricardo Rom\'an-Ancheyta\,\orcidlink{0000-0001-6718-8587}}
\email{ancheyta@fata.unam.mx}
\affiliation{Centro de F\'isica Aplicada y Tecnolog\'ia Avanzada, 
Universidad Nacional Aut\'onoma de M\'exico, \\
Boulevard Juriquilla 3001, 76230 Quer\'etaro, Mexico}
 
\date{\today}
\begin{abstract} 
We obtain time-resolved spectra of spontaneous emission and resonance 
fluorescence of a single multilevel emitter where two antiparallel transitions 
interfere and cause quantum beats. After rising as a single broad peak, the 
spontaneous emission spectrum turns into a doublet of subnatural peaks 
and then fades for long times. For strong field resonance 
fluorescence, the beat signature is the formation of doublet sidebands, 
which initially grow asymmetrically but end up symmetrical. We stress 
the \textcolor{black}{filter bandwidth's crucial role} in the spectral 
resolution and causal evolution. 
\end{abstract}

\maketitle

\section{Introduction} 
Quantum beats (QB), the modulation of the fluorescence signal of a multilevel 
system by the interference of two or more close frequencies of transitions 
from, e.g., the fine or hyperfine structure, are among the more familiar and 
interesting manifestations of quantum mechanics~\cite{Breit}, with applications 
in high-resolution spectroscopy and quantum technologies \cite{Haroche,Zajonc,Hack91,Havey+14,LaForge+20,NOBC10,CSP+13,
Trifonov20,Cai+23,Lee+23,WuLiWu23}. 
Its observation requires ignoring the frequencies of the emitted photons 
when observed by a broadband detector after the atom (or other emitter) 
was prepared in a superposition of states. In most quantum beat 
experiments, light from spontaneous emission is observed, but it can also 
be from resonance fluorescence after cw laser excitation \cite{CSGA24} or 
other processes \cite{FiSw05,Yano+09}. 

A QB is a time-resolved effect, so its spectrum, the frequencies that 
compose the beats, is obtained from the Fourier transform of the excited 
state populations if spontaneous emission is measured or from the stationary 
dipole correlation function, the Wiener-Khintchine (WK) formula, for resonance 
fluorescence. Besides being defined differently, these spectra are only 
snapshots obtained long after the start of the interaction. They do not tell 
the story of how they reached such spectra. This is problematic for general 
time-dependent excitation (pulsed, chirped, sudden, constant, etc.) or 
internal structures, like the one causing the QBs we study here, 
and other couplings of the emitter. 

To address the need for a comprehensive definition of time-dependent spectra 
(TDS), Eberly and W\'odkiewicz (EW) devised the Physical Spectrum 
\cite{EbWo77}, based on photon counting at the relevant frequencies 
\cite{MandelWolf} and filtering due to the detector's bandwidth \cite{BornWolf}, 
intrinsic components of the measurement process that guarantee positivity 
and compliance with the time-energy uncertainty, acknowledging that one 
cannot resolve both spectrum and time development arbitrarily at the same 
time. 

In this \textcolor{black}{paper} we show theoretically that quantum beat 
effects can be nicely captured in the EW TDS of spontaneous emission 
and resonance fluorescence from a single two-level atom with angular 
momentum $J=1/2 -J'=1/2$, subject to Zeeman splittings, the beats 
resulting from interference among the $\pi$ transitions ($m=m'$) 
\cite{CSGA24}. 

Interestingly, the time-resolved spontaneous emission spectrum develops 
from a broad single peak into a doublet, the signature of beats, and then 
fades away. This fading sounds odd but, after all, there is only one photon 
to be emitted, after which there is no light and thus no spectrum. And, 
very  importantly, we obtain and provide an exact analytical result. 

For stationary resonance fluorescence, if the laser and magnetic fields 
are strong, the beats have a well-defined mean and modulation frequency 
in observables such as the intensity and two-time correlation functions 
\cite{CSGA24}. The dipole correlation gives way to a Wiener-Khintchine, 
Mollow-like spectrum given by a central peak and sidebands that are 
actually doublets, again, the signature of QBs. \textcolor{black}{In contrast, 
we show that the time-dependent spectrum evolves asymmetrically with 
oscillations at the spectral components' frequencies and ends} up symmetric. 

A Fabry-Perot type of filter has a crucial role in the EW spectra: for once, 
it selects the frequency component of the emission \cite{BornWolf}, but it 
also blurs the order of photon arrivals at the detector, which manifests as 
linewidth narrowing in the TDS of spontaneous emission \cite{EbKW80}. 
As a consequence, the bandwidth must be chosen so that the doublets 
are sufficiently resolved, but also that the beating wave packets are not 
entirely bypassed. 

Nonstationary systems have been common subjects of EW TDS. Ideal 
optical cavities \cite{SaNE83,CaCS96,YLYO22}, anisotropic polaritons
~\cite{RRA21}, quantum thermometers~\cite{RRA20}, superconducting 
qubits~\cite{SSRA22}, and oscillating mirrors~\cite{GHD+10,Mirza15} 
produce long-term stabilized spectra. Turn-on \cite{EbKW80} and turn-off 
\cite{HuTE82} of the atom-laser interaction lead to the birth and death 
(by spontaneous emission), respectively, of the Mollow triplet spectrum 
of resonance fluorescence \cite{Mollow69}. Even under continuous wave 
driving, some fluorescence processes have nontrivial TDS under specific 
conditions, such as initial coherence \cite{GoMo87} and coherent 
population trapping \cite{JLDS89,MiJa24}. Recently, with others, we 
studied the TDS of blinking resonance fluorescence \cite{RSHC18}, 
observing that the Mollow spectrum emerges well before the ultra-narrow 
peak due to electron shelving \cite{BuTa00,EvKe02,CaRG16}. TDS of 
light~\cite{EbWo77} are of practical interest for emerging applications in 
quantum optics \cite{vEnk17,HoFi10} and quantum control 
\cite{Muga19,Ancheyta23}, where measurements are constrained by time. 

\section{Model} 
The studied system, depicted in Fig.~\ref{spectrumSEfull}~(a), consists 
of a single two-level emitter with angular momentum states $|J,m\rangle$, 
where $J=1/2$ for both levels and  $m=\pm J$ is the magnetic quantum 
number, giving the upper states $|1\rangle \equiv |1/2, -1/2 \rangle$ and 
$|2\rangle \equiv |1/2, 1/2 \rangle$, and the lower states 
$|3\rangle \equiv |1/2, -1/2 \rangle$ and $|4\rangle \equiv |1/2, 1/2 \rangle$. 
This level configuration is found in $^{198} \mathrm{Hg}^+$ \cite{PoSc76} 
and charged quantum dots \cite{Steel05}, for example. We only observe 
the $\pi$ transitions ($m=m'$), which have dipole matrix elements 
$\mathbf{d}_1 = \langle 1| \hat{\mathbf{d}} |3 \rangle 
	= - \mathcal{D} \mathbf{e}_z /\sqrt{3}$, and 
$\mathbf{d}_2 = \langle 2| \hat{\mathbf{d}} |4 \rangle  = - \mathbf{d}_1$, 
where $\mathcal{D}$ is the reduced dipole matrix element and the sign 
difference implies that they are antiparallel. The $\pi$ transitions couple 
to linearly polarized light along the quantization axis $z$. 

We remove level degeneracies by the application of a static magnetic field 
$B_z$ along the $z$ direction, the Zeeman effect, with splittings 
$g_u \mu_B B_z$ and $g_\ell \mu_B B_z$ for the upper and lower levels, 
where $g_u$ and $g_\ell$ are the respective Land\'e $g$-factors, and 
$\mu_B$ is Bohr's magneton. Thus, the difference Zeeman splitting is 
$\delta = (g_u -g_{\ell}) \mu_B B_z /\hbar $, so the effective transition 
frequencies of the $\pi$ transitions are $\omega_0 = \omega_{13}$ and 
$\omega_{24} = \omega_{13} +\delta$. 

The atom is excited by a monochromatic laser field of frequency $\omega_L$, 
wave vector $k_L$, and amplitude $E_0$, linearly polarized in the $z$ 
direction, propagating in the $x$ direction, $\mathbf{E}_L (x,t) 
= E_0 e^{i (\omega_L t - k_L x)} \mathbf{e}_z 	+\mathrm{c.c.}$, 
driving only the $\pi$ transitions. The Rabi frequency is then 
$\Omega = E_0 \mathcal{D} / \sqrt{3}\,\hbar$. The atom-laser detuning is 
referenced to the $|1 \rangle - |3 \rangle$ transition, that is,  
$\Delta = \omega_L - \omega_{13}$, so that $\Delta - \delta$ is the detuning 
on the $|2 \rangle - |4 \rangle$ transition. 
 
Defining the atomic operators $A_{jk} = |j \rangle \langle k|$, where $j \neq k$ 
denotes coherences, and $j=k$ denotes populations, and in the frame rotating 
at the laser frequency, we finally have the Hamiltonian \cite{KiEK06b}
\begin{eqnarray} 	\label{eq:Hamiltonian}
H &=& - \hbar \Delta A_{11} - \hbar (\Delta -\delta) A_{22}  
	+\hbar B_z (A_{22} +A_{44}) \nonumber	\\ 
&& + \hbar \Omega \left[ (A_{13} -A_{24})  + \mathrm{h.c.} \right] .
\end{eqnarray}

The excited states decay by spontaneous emission either in the $\pi$ 
transitions, emitting photons with linear polarization at rate $\gamma_{\pi}$, 
or in the $\sigma$ transitions, emitting photons with circular polarization at 
rate $\gamma_{\sigma}$, with branching ratios of ${1}/{3}$ and ${2}/{3}$, 
respectively \cite{KiEK06b}. The total decay rate of each excited state is 
then $\gamma=\gamma_{\pi} +\gamma_{\sigma}$.  

The positive-frequency part of the source field operator in the 
far-field zone is given by \cite{KiEK06b,CSGA24} 
\begin{eqnarray}  	\label{eq:field-atomOps}
\hat{E}_{\pi}^+ (\mathbf{r}, t) &=&  f_{\pi}(r)  
	\left[ A_{31} (t) - A_{42} (t) \right] \mathbf{e}_z , 
\end{eqnarray}
where $f_{\pi}(r)= -\omega_0^2 \mathcal{D}/4\sqrt{3}r \pi \epsilon_0 c^2$, 
is a constant factor we henceforth drop, so the total intensity is given 
by the expectation values of the excited state populations,  
\begin{equation} 	\label{eq:time_Intensity_pi}
    I_{\pi} (\mathbf{r}, t) = \langle \hat{E}_{\pi}^- (\mathbf{r}, t) 
	\hat{E}_{\pi}^+ (\mathbf{r}, t) \rangle
= \langle A_{11} (t) \rangle +\langle A_{22} (t) \rangle  .
\end{equation}

\section{Time-Dependent Spectra} 
We calculate the TDS of the $\pi$ transitions using the physical spectrum 
by Eberly and W\'odkiewicz \cite{EbWo77,Carm08}, defined as  
\begin{equation}\label{eq:EWspec}
\begin{aligned}
S(\nu,t,\Gamma) =&\, \Gamma \int_{t_0}^t dt_1  \int_{t_0}^t dt_2 \,\, 
	e^{-(\Gamma/2 -i \nu)(t-t_1)} \\ 
 	 &\times	e^{-(\Gamma/2 +i \nu)(t-t_2)} 
	\langle \hat{E}_{\pi}^- (t_1) \hat{E}_{\pi}^+ (t_2) \rangle  \,, 
\end{aligned}
\end{equation}
where $t$ is the time elapsed since the laser turn-on $t_0=0$, 
$t_2 -t_1 \equiv \tau \geq 0$ is the time delay between two measurements of 
the field correlation, $\nu=\omega -\omega_L$ is the detuning of the laser 
frequency $\omega_L$ from the filter's frequency $\omega$, $\Gamma$ 
is the filter's bandwidth, and  
$\langle \hat{E}_{\pi}^- (t_1) \hat{E}_{\pi}^+ (t_2) \rangle$ is the dipole 
correlation function. Inserting Eq.~(\ref{eq:field-atomOps}) into 
Eq.~(\ref{eq:EWspec}), noting that 
$\langle A_{13}(t_1) A_{42}(t_1+\tau) \rangle$ and  
$\langle A_{24}(t_1) A_{31}(t_1+\tau) \rangle$ are zero due to initial 
conditions, we have the spectrum written in terms of the atomic operators, 
\begin{eqnarray}  \label{eq:spectra-pi} 
\lefteqn{
S(\nu, t, \Gamma) = 2\Gamma \mathrm{Re}  \int_{0}^t dt_1  \, 
	e^{-\Gamma(t-t_1)} \int_{0}^{t-t_1} d\tau  \, e^{(\Gamma/2 -i \nu) \tau}  } 
	\quad \nonumber \\
  &&  \times \left[ \langle A_{13} (t_1) A_{31} (t_1+\tau) \rangle 
	+ \langle A_{24} (t_1) A_{42} (t_1+\tau)  \rangle \right]    .
\end{eqnarray}

For a stationary process ($t\to \infty\!$) and unrealistic perfect resolution 
($\Gamma\to0$), we recover the Wiener-Khintchine (WK) spectrum, $S_{\mathrm{WK}}(\nu) = \mathrm{Re}  \int_{0}^{\infty} d\tau  
e^{-i \nu\tau}$ $\langle \hat{E}_{\pi}^- (0) \hat{E}_{\pi}^+ (\tau) \rangle$ 
\cite{EbWo77,BornWolf}. Often a finite detection bandwidth is added phenomenologically, with the replacement $\nu \to \nu+i\Gamma$ 
\cite{KiEK06b}. The WK spectrum implies a long measurement time, 
missing whatever changes a spectrum has undergone. 
\begin{figure}[t]
\includegraphics[width=8.5cm,height=6.25cm]{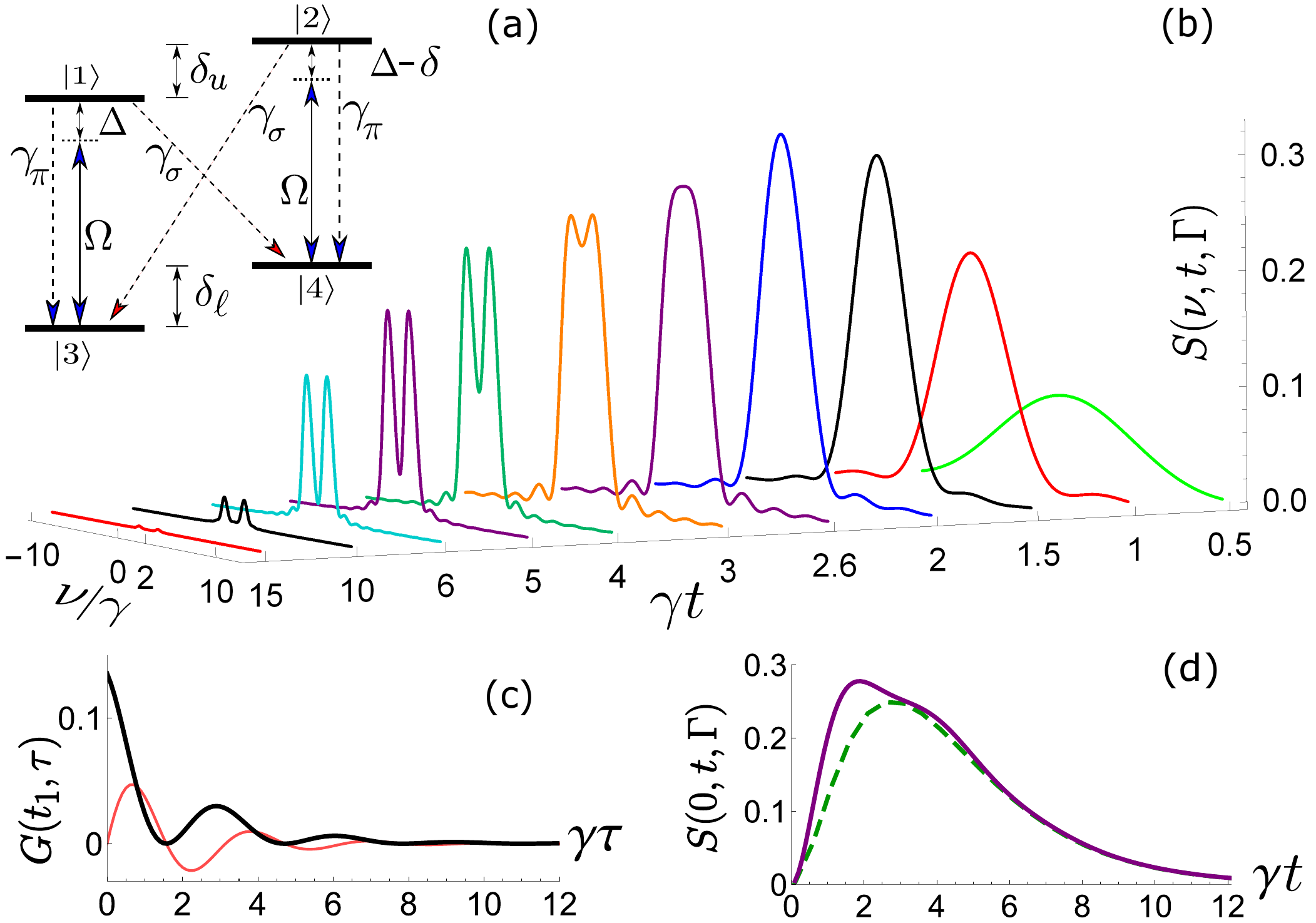} 
\caption{\label{spectrumSEfull} 
(a) Atom-laser system (see main text for details). (b) TDS 
$S(\nu,t,\Gamma)$ of spontaneous emission, where 
$\Omega$=$\Delta$=0, for $\delta = -2\gamma$, $\Gamma= \gamma/2$, 
and initial condition 
$\langle A_{11}(0) \rangle = \langle A_{22}(0) \rangle =1/2$, with the other 
$\langle A_{jk}(0) \rangle =0$.  (c) Real (black-thick) and imaginary 
(red-thin) parts of the correlation $G(t_1 =2/\gamma, \tau)$. (d) TDS at 
$\nu= \omega -\omega_{13}=0$ for $\delta= -2\gamma$ (solid) and 
$\delta=0$ (dotted, half the actual amplitude).}
\end{figure}

Now, we focus on our results, presenting the technicalities of the analytical 
and numerical calculations in Appendix~\ref{App_ME}. For QBs to be observable in the
system, the primary conditions are that we break the level degeneracy, 
$\delta \neq 0$, and set the initial state in a superposition of both excited 
states for spontaneous emission or both ground states for resonance 
fluorescence \cite{CSGA24}.  

A major result of this \textcolor{black}{work} is that we obtained the EW TDS of spontaneous 
emission analytically, 
\begin{eqnarray}        \label{eq:S_SE}
S_{\mathrm{SE}}(\nu, t, \Gamma) 
= \frac{F(t,0)}{(\gamma -\Gamma )^2 +4\nu^2} 
    +\frac{F(t,\delta)}{(\gamma -\Gamma )^2+4 (\nu +\delta)^2} ,
    \nonumber \\
F(t,x) \equiv 2 \Gamma    
    \left( e^{-\Gamma t} +e^{-\gamma t} \right) 
	 -4 \Gamma e^{-\frac{1}{2} (\gamma +\Gamma) t } 
    \cos ((\nu +x) t) .  \nonumber \\
\end{eqnarray}
Figure~\ref{spectrumSEfull}(b) shows a sequence of traces of the 
time-dependent spontaneous emission spectrum. Initially, for times 
$t \sim 1/\Gamma$, the spectrum is not fully developed, given by a single 
peak; this is in the filter's filling time. Then, the spectrum splits and 
becomes a doublet: the peak at $\nu=0$ comes from the transition 
$|1\rangle\rightarrow|3\rangle$, while the peak at $\nu=-\delta=2$ comes 
from the transition $|2\rangle\rightarrow|4\rangle$. The separation 
$|\delta|$ is the modulation (beat) frequency of the decay of the correlation  
$G(t_1,\tau) = \langle \hat{E}_{\pi}^- (t_1) \hat{E}_{\pi}^+ (t_1+\tau) \rangle$;  
see Fig.~\ref{spectrumSEfull}(c). In the degenerate case ($\delta=0$), 
the spectrum Eq.~(\ref{eq:S_SE}), lacking quantum beats, is reduced to a 
single peak \cite{EbKW80}, which grows more slowly than any peak of the 
doublet, as seen in Fig.~\ref{spectrumSEfull}(d). But what might seem 
surprising is that the doublet fades for times $t \gg 1/\Gamma$. The 
explanation is simple: the atom emits only one photon. At long times, 
it is very likely that the atom has already done so, and, eventually, there 
would be no light in the filter to be transmitted to the detector. 

We have chosen a filter bandwidth, $\Gamma = |\delta|/4$, narrow enough 
to resolve the doublet but wide enough to fill the filter rapidly and allow for 
the spectrum to reveal its shape not long after $t \sim \Gamma^{-1}$,  
a manifestation of the time-energy uncertainty. With a very narrow 
bandwidth, $\Gamma \ll |\delta|$, it would take a long time to reveal the 
final spectral shape, not capturing the QB signature. A large bandwidth, 
$\Gamma > |\delta|$, would give, if anything, a short-lived single-peak 
spectrum; the broadband regime $\Gamma\gg |\delta|$ is that of 
time-resolved beats. Thus, it takes time to measure a spectrum, but our 
goal is to capture spectral transients \textcolor{black}{by taking only a finite 
time} for the measurement. 

Another remarkable effect is the subnatural width of the peaks, 
$\gamma-\Gamma$, a signature of interference of the source light with 
the filter, which causes uncertainty in the time of arrival and transmission 
of photons. In the extreme narrowing case, $\Gamma=\gamma$, the 
spectrum Eq.~(\ref{eq:S_SE}) is reduced to 
$S_{\rm SE}(\nu,t,\Gamma)\!=\!\frac{1}{2}\Gamma t^2e^{-\Gamma t}
\{{\rm sinc}^2(\nu t/2)\!+\!{\rm sinc}^2[(\nu+\delta)t/2]\}$, where 
${\rm sinc}(z)\equiv\sin(z)/z$. The peak frequencies ($\nu=0$ and 
$\nu=-\delta$) grow and die as $\frac{1}{2}\Gamma t^2 e^{-\Gamma t}$, 
while their widths are proportional to $t^{-1}$, i.e., for $\Gamma=\gamma$, 
$S_{\rm SE}(\nu,t,\Gamma)$ evolves into two Dirac delta functions but with 
vanishing amplitude. The lateral lobes (zeros) of ${\rm sinc}(z)$ explain the 
small bottom oscillations in the TDS of Fig.~\ref{spectrumSEfull}(b). 

Recall that spontaneous emission is a transient process, so it does not 
have a Wiener-Khinchine spectrum. Conveniently, its spectrum has been 
defined as the Fourier transform of the upper state population, not of the 
dipole correlation. A more formal approach \cite{Carm02} is to do a double 
integral of the dipole correlation, 
$P(\nu) \propto \int_0^T dt_1 \int_0^T dt_2 \, e^{-i\nu (t_1-t_2)} G(t_1,t_2)$,
still with perfect resolution, obtaining the expected two Lorentzians 
centered at $\nu=0$ and $\nu=-\delta$ in the long-time limit. These 
definitions may be useful but miss entirely spontaneous emission's 
intrinsic transient, single-photon nature.

\begin{figure}[b]
\includegraphics[scale=0.24]{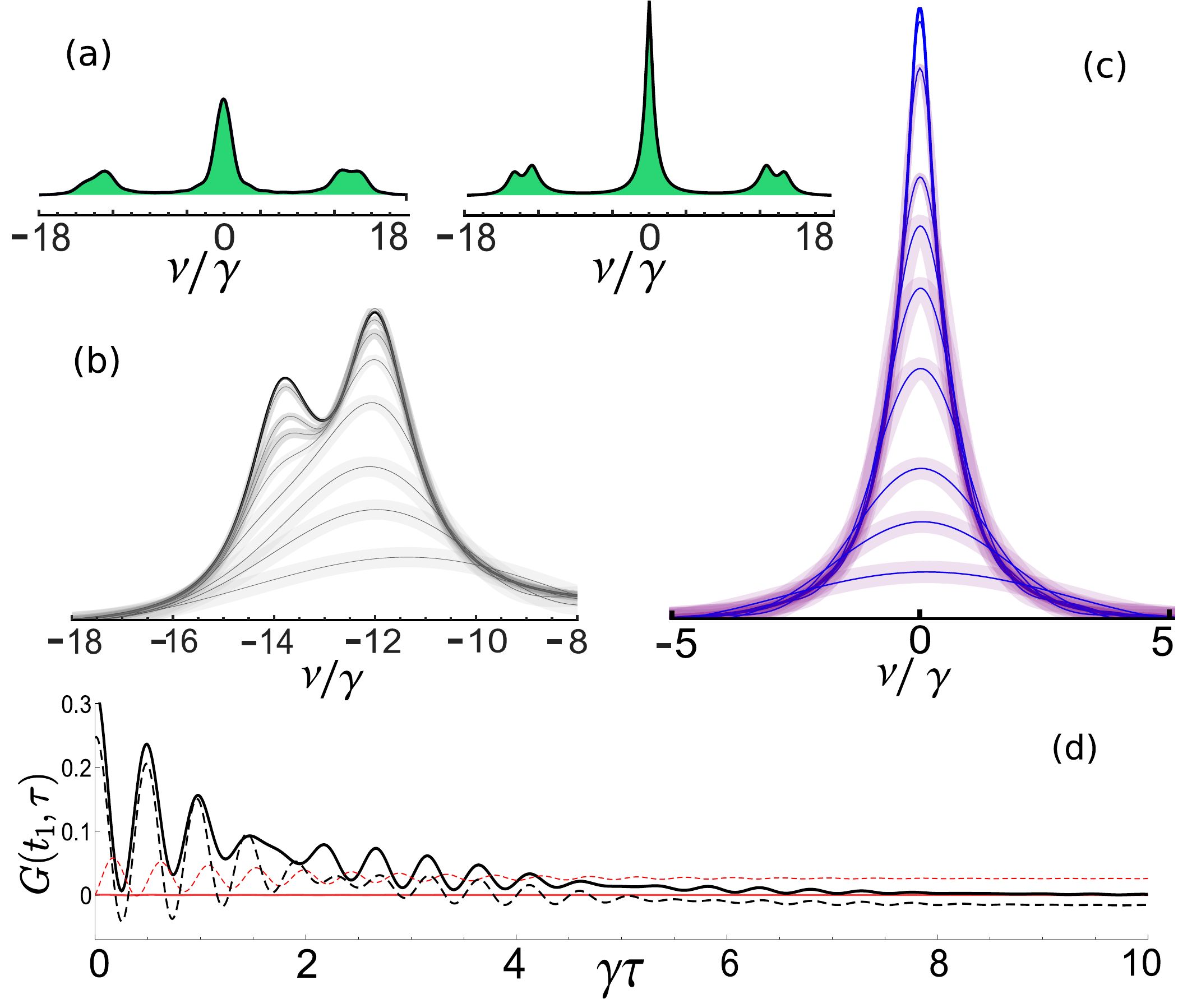} 
\caption{\label{fig_full-left-ctr-G} 
(a) Full spectrum $S(\nu,t,\Gamma)$ at short, $\gamma t=4$ (left), and long 
$\gamma t =15$ (right), times. Growth of the left sideband (b) and central 
band (c) at the increasing measurement times 
$\gamma t =1, {3}/{2}, 2, 3, 4, 5, 6, 10, 15, 20$ (lower to higher traces). 
(d) Dipole correlation $G(t_1,\tau)$ for $\gamma t_1=1$ (dashed) and 
$\gamma t_1 =7$ (solid), real part (thick) and imaginary (thin). The other 
parameters are: $\Omega = 6\gamma$, $\delta = -7\gamma$, $\Delta =0$, 
and $\Gamma= 0.5\gamma\simeq \Omega_{\rm beat}/2$. The initial 
condition is $\langle A_{33}(0) \rangle = \langle A_{44}(0) \rangle =1/2$, 
and the other $\langle A_{jk}(0) \rangle =0$. }
\end{figure}
Now we consider the contrasting case of resonance fluorescence by a 
highly  nondegenerate atomic system driven by a strong laser field, 
$\Omega \sim \delta \gg \gamma$ (we choose, for simplicity, resonant 
excitation, $\Delta=0$). Fig.~\ref{fig_full-left-ctr-G}(a) shows the full 
spectrum at a short time  $\gamma t=4$ (left), when it has not fully 
developed yet, and at a long time $\gamma t=15$ (right), very near its 
stationary shape. It is a Mollow-type spectrum with sidebands that are 
doublets with peaks of unequal heights at frequencies 
\cite{KiEK06b,CSGA24} 
\begin{eqnarray} 	\label{eq:Rabis} 
\Omega_1 &=& \sqrt{ 4\Omega^2 +\Delta^2 }  , \qquad 
	\Omega_2 = \sqrt{ 4\Omega^2 +(\delta -\Delta)^2 }  . \hspace{5mm}
\end{eqnarray}

\begin{figure}[t] 
\includegraphics[scale=0.43]{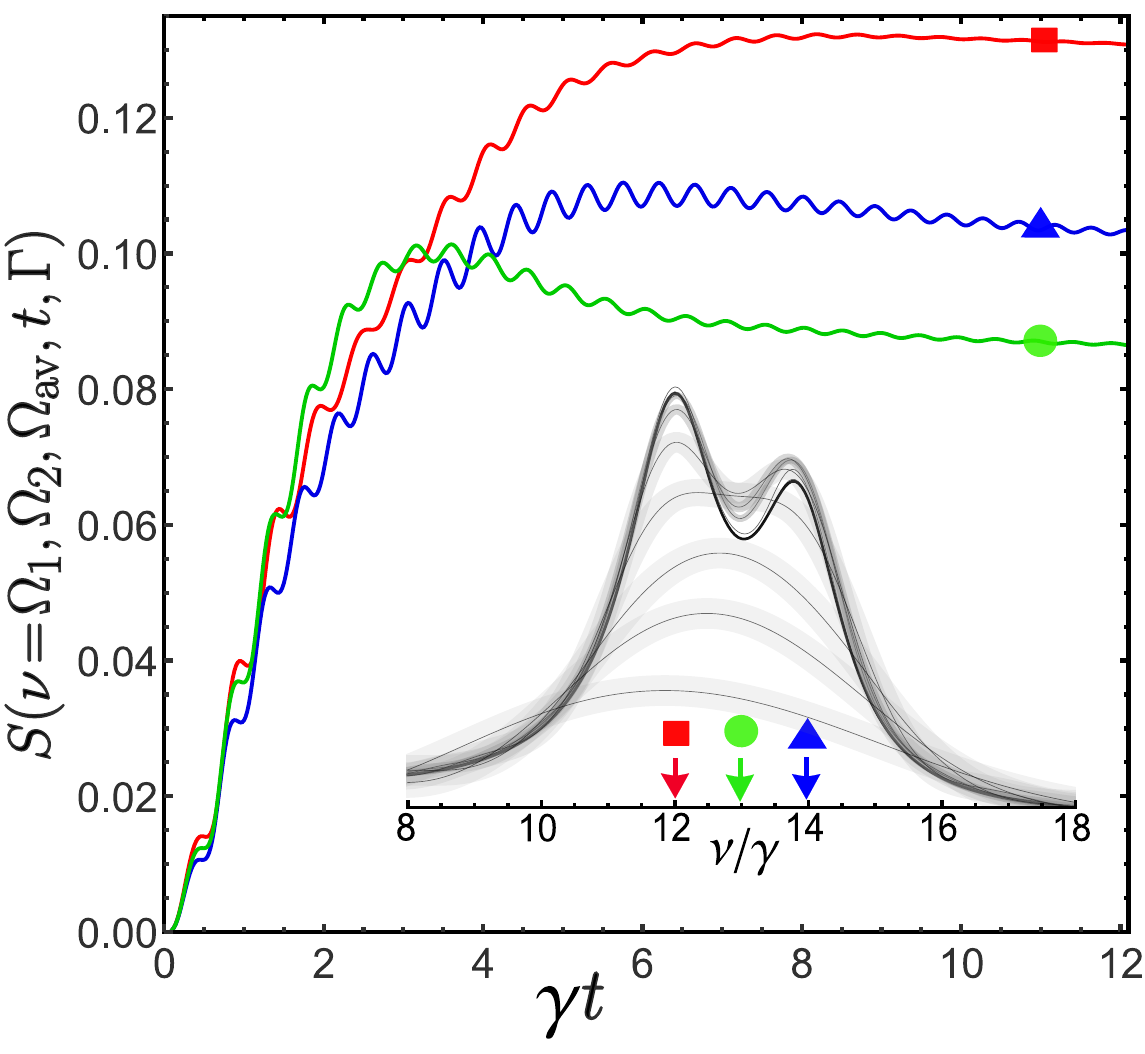} 
\caption{\label{fig_right_sb} 
Evolution of the right doublet sideband of $S(\nu,t,\Gamma)$ (gray lines) 
and at the frequencies of the peaks at $\nu=\Omega_1$ (square) and  
$\nu=\Omega_2$ (triangle), and of the dip at $\nu=\Omega_{\mathrm{av}}$ 
(circle). Other labeling and parameters are as in Fig.~\ref{fig_full-left-ctr-G}. }
\end{figure}

We zoom in on the evolution of the left sideband and central band in 
Figs.~\ref{fig_full-left-ctr-G}(b,c), and of the right sideband in the inset of 
Fig.~\ref{fig_right_sb}, from an early measurement time, $\gamma t =1$, 
up to a long time, $\gamma t =20$. Comparing the right and left sidebands, 
we see that the spectrum evolves strongly asymmetrically but eventually 
reaches a symmetric shape. The transient asymmetry is due to nonzero 
detunings, here only $\delta$. [Fig.~4 of Ref.~\cite{RSHC18} shows a similar 
transient asymmetry.] The peaks at $|\Omega_1|=12\gamma$ stabilize a 
little sooner than those at $|\Omega_2|\simeq 13.9\gamma$ because the 
$|1\rangle -|3\rangle$ transition frequency is closer to the laser frequency. 
The small oscillations in the evolution at the frequencies of the right sideband 
peaks and the dip, Fig.~\ref{fig_right_sb}, missing in the steady state, are 
also due to the detuning. 

The features of the spectrum are, of course, related to the properties of the 
source dipole correlation $G(t_1,\tau)$. Most important is that the doublet 
sidebands result from beats with well-defined mean, 
$\Omega_{\mathrm{av}} = \frac{1}{2}(\Omega_2 +\Omega_1)$, and 
modulation frequency, 
$\Omega_{\mathrm{beat}} = \frac{1}{2}(\Omega_2 -\Omega_1)$. 
These beats occur only in the nondegenerate case ($\delta \neq 0$), 
where the two $\pi$ transitions are unbalanced, evolving with Rabi 
frequencies $\Omega_1$ and $\Omega_2$ \cite{CSGA24}. 
Fig.~\ref{fig_full-left-ctr-G}(d) shows $G(t_1,\tau)$ for $\gamma t_1=1$, 
which generates the lowest (first trace) spectrum, and $\gamma t_1=7$, 
a time long enough to see any further changes and the doublets are 
already well-formed. Detunings give the correlation a transient nonzero 
imaginary part (when the spectrum is asymmetric) that vanishes in the 
steady state. 

Again, our choice of the filter bandwidth, 
$\Gamma=\gamma/2\simeq \Omega_{\rm beat}/2$, is made on the 
grounds of not only resolving the doublet, 
$\Gamma < \Omega_2 -\Omega_1$, but also capturing the sideband 
oscillations and beat modulation in $G(t_1,\tau)$ at times 
$t \sim \Gamma^{-1}$. With a narrower $\Gamma$, not only do we miss 
the beat and make the oscillations smaller, but the whole spectrum would 
evolve more symmetrically and monotonically (See Fig.~\ref{Four_bands} 
in Appendix~\ref{App_Filter}). Causality would prevent the formation of the doublet before 
the beat occurs. We cannot measure quantum interference before 
interference occurs. This effect is analogous to the nonobservation of the 
Rabi doublet in cavity QED before a Rabi oscillation is completed 
\cite{YLYO22}. 

The nondegenerate $J=1/2 - J'=1/2$ system is relatively simple yet 
nontrivial that shows quantum interference to several orders of coherence 
functions. Because the $\pi$ transitions are antiparallel, that is, they do 
not end nor start in common states, one would not expect quantum beats 
in the intensity; they have to be looked at in fluctuations 
\cite{CSGA24,Zajonc,NOBC10,CSP+13,FiSw05,Yano+09}, like the dipole 
correlation in this work. Still, there are beats in the intensity of resonance 
fluorescence. The point is that there is an underlying coherence in the 
scattering due to the initial state preparation for spontaneous emission 
and the laser for resonance fluorescence, which allows for interference in 
higher-order functions such as intensity-intensity and 
intensity-amplitude correlations \cite{CSGA24}. 

An interesting application of the EW Physical Spectrum consists of 
correlating photons of selected spectrum components  
\cite{Nienhuis93a,Nienhuis93b}. But this comes with the difficulty of 
handling sets of time-ordered integrals for every photon sequence. There 
are theoretical and experimental efforts to get around the complications 
due to filtering (see the review \cite{dV+L20} and also 
\cite{NgPC24,CBChSM98}), especially for resonance fluorescence in 
cavity QED. It would be very interesting to study frequency-selected photon 
correlations from the spectrum of quantum beats. However, this is beyond 
the purpose of this \textcolor{black}{work}.

\section{Conclusions} 
We have calculated time-dependent spectra of spontaneous emission and 
resonance fluorescence (pillars of quantum optics) of a system that features 
quantum beats using the Eberly-W\'odkiewicz physical spectrum. The 
resolution of temporal features in the spectra crucially depends on the 
bandwidth of the detection technique. Our results show that the richness 
of behavior of time-dependent spectra can be revealed beyond what is 
possible with the Wiener-Khintchine spectrum. With this work, we hope 
to stimulate further studies of time-resolved spectra for stationary and 
nonstationary systems for a variety of applications, and fundamental 
questions in quantum mechanics, such as causality and complementarity.

\section{Acknowledgments.} 
H.M.C.-B. thanks Programa de Movilidad Nacional UAEM-UNAM, Convenio 
No. 42467-2177-8-IX-15, 
for support to perform part of his work in CFATA. R.R.-A. thanks 
DGAPA-UNAM, M\'exico for support under Project No. IA104624.

\appendix

\section{Master Equation and Quantum Regression Formula} 
\label{App_ME}
The dynamics of the atom-laser-reservoir system are described by the 
master equation for the reduced atomic density operator, $\rho$. In a 
frame rotating at the laser frequency this is $\partial_t \tilde{\rho} 
 = -\frac{i}{\hbar} [H,\tilde{\rho}] +\mathcal{L}_{\gamma}  \tilde{\rho} $, 
where $-\frac{i}{\hbar} [H,\tilde{\rho}]$ describes the coherent atom-laser 
interaction and $\mathcal{L}_{\gamma}  \tilde{\rho}$ the damping due to 
spontaneous emission \cite{KiEK06b,Agarwal74}, 
$\mathcal{L}_{\gamma} \tilde{\rho} 
	= \frac{1}{2} \sum_{i,j=1}^2 \gamma_{ij} \left( 2S_i^- \tilde{\rho} S_j^+ 
	- S_i^+ S_j^- \tilde{\rho} -\tilde{\rho} S_i^+ S_j^-  \right) 	
    + \frac{\gamma_{\sigma}}{2} \sum_{i=3}^4 \left( 2S_i^- \tilde{\rho} S_i^+ 
	- S_i^+ S_i^- \tilde{\rho} -\tilde{\rho} S_i^+ S_i^-  \right)$,	
where $S_1^- = A_{31}$, $S_2^- = A_{42}$, $S_3^- = A_{32}$, and 
$S_4^- = A_{41}$. 
Noting that the coherences for the $|1\rangle -|4\rangle$, 
$|2\rangle -|3\rangle$, $|1\rangle -|2\rangle$, and $|3\rangle -|4\rangle$ 
transitions vanish at all times thus, instead of 16 equations, we are left 
with a set of 8 relevant homogeneous equations \cite{CSGA24}: 
\begin{eqnarray} 	\label{eq:BlochEqs1}
\langle \dot{A}_{11} \rangle &=& - \gamma \langle A_{11} \rangle 
	+i \Omega (\langle A_{31} \rangle -\langle A_{13} \rangle) 	, 
	\nonumber \\ 
\langle \dot{A}_{13} \rangle &=& -\left( \frac{\gamma}{2} 
	+i\Delta  \right) \langle A_{13} \rangle - i \Omega (\langle A_{11} \rangle 
	- \langle A_{33} \rangle) 	,  \nonumber \\
\langle \dot{A}_{22} \rangle &=& -\gamma \langle A_{22} \rangle 
	- i \Omega (\langle A_{42} \rangle - \langle A_{24} \rangle) , 
	\nonumber \\ 
\langle \dot{A}_{24} \rangle &=& -\left( \frac{\gamma}{2} 
	+i (\Delta -\delta) \right) \langle A_{24} \rangle 
	+ i \Omega (\langle A_{22} \rangle - \langle A_{44} \rangle)  , 
	\nonumber \\ 
\langle \dot{A}_{31} \rangle &=& -\left( \frac{\gamma}{2} 
	-i\Delta  \right) \langle A_{31} \rangle + i \Omega (\langle A_{11} \rangle 
	- \langle A_{33} \rangle) 	, \nonumber \\ 
\langle \dot{A}_{33} \rangle &=& \gamma_1 \langle A_{11} \rangle 
	+\gamma_{\sigma} \langle A_{22} \rangle 
	-i \Omega (\langle A_{31} \rangle - \langle A_{13} \rangle) 	, 
	\nonumber \\
\langle \dot{A}_{42} \rangle &=& -\left( \frac{\gamma}{2} 
	-i (\Delta -\delta) \right) \langle A_{42} \rangle 
	- i \Omega (\langle A_{22} \rangle - \langle A_{44} \rangle)   ,  
	\nonumber \\ 
\langle \dot{A}_{44} \rangle &=& \gamma_{\sigma}  \langle A_{11} \rangle    
	+\gamma_2 \langle A_{22} \rangle 
	+i \Omega (\langle A_{42} \rangle - \langle A_{24} \rangle) . 
\end{eqnarray}
This then lets us define a simpler Bloch vector 
$\mathbf{Q} \equiv \left( A_{11}, A_{13}, A_{22}, A_{24}, 
A_{31}, A_{33}, A_{42}, A_{44} \right)^T$. A set of homogeneous 
equations translates into a great reduction in numerical calculations 
compared to the inhomogeneous case, obtained by eliminating, e.g., 
the population $\langle A_{44} \rangle$ via the conservation of probability. 
The resulting Bloch equations for the atomic expectation values can be 
written compactly as $\partial_t \langle \mathbf{Q}(t) \rangle 
	= \mathbf{M} \langle \mathbf{Q}(t) \rangle$, where $\mathbf{M}$ 
is an $8 \times 8$ matrix of coefficients~\cite{CSGA24} and the formal 
solution is given by $\langle \mathbf{Q}(t) \rangle 
= e^{\mathbf{M} t} \langle \mathbf{Q}(0) \rangle$.

The advantage of solving only eight homogeneous equations carries over 
to the two-time correlations. For this, we apply the quantum regression 
formula \cite{Carm02}, that is, 
$\partial_\tau \langle \mathbf{W}(t_1,\tau) \rangle 
	= \mathbf{M} \langle \mathbf{W} (t_1,\tau) \rangle$, 
where we defined a Bloch-like vector 
$\mathbf{W} (t_1,\tau)  = A_{jk}(t_1) \mathbf{Q}(t_1+\tau)$, and has 
the formal solution $\langle \mathbf{W} (t_1,\tau) \rangle 
	= e^{\mathbf{M} \tau} \langle  \mathbf{W} (t_1,0) \rangle$.  
We define the two-time operator functions 
$\mathbf{U} (t_1,\tau) = A_{13} (t_1) \mathbf{Q} (t_1+\tau)$ and 
$\mathbf{V} (t_1,\tau) = A_{24} (t_1) \mathbf{Q} (t_1+\tau)$,
 from which the second and seventh terms, respectively, are the 
 correlations given in the second line of Eq.~(\ref{eq:spectra-pi}).
Their expectation values have initial conditions (at $\tau =0$): 
\begin{subequations} 	\label{eq:U_initial}
\begin{eqnarray} 
\langle \mathbf{U} (t_1,0) \rangle 
	&=& \left( 0, 0, 0, 0, \langle A_{11} (t_1) \rangle,  
	 \langle A_{13} (t_1) \rangle, 0, 0 \right)^T , \quad  \\
\langle \mathbf{V} (t_1,0) \rangle 
	&=& \left( 0, 0, 0, 0, 0, 0, 
	\langle A_{22} (t_1) \rangle, 
	  \langle A_{24} (t_1) \rangle  \right)^T . \quad 
\end{eqnarray}
\end{subequations}

To observe beats in spontaneous emission, $\Omega=\Delta=0$, we need 
a nonzero difference of Zeeman detunings, $\delta \neq0$, and both upper 
states must be initially nonzero, ideally equal \cite{CSGA24}, so the initial 
Bloch vector is 
\begin{eqnarray} 	\label{eq:initialCondBeatsSpEm}
\langle \mathbf{Q}(0) \rangle_{\mathrm{SE}} 
	&=& \left( 1/2, 0, 1/2, 0, 0, 0, 0, 0  \right)^T .  
\end{eqnarray}
In this case, we solve the Bloch equations (\ref{eq:BlochEqs1}) 
analytically: $\langle A_{11}(t) \rangle 
= \langle A_{22}(t) \rangle = \frac{1}{2} e^{-\gamma t}$, 
$\langle A_{33}(t) \rangle = \langle A_{44}(t) \rangle 
= \frac{1}{2}(1-e^{-\gamma t})$, with vanishing coherences. Note that the 
intensity, $I_{\pi}^{SE} =e^{-\gamma t}$ [Eq.~(\ref{eq:time_Intensity_pi})], 
shows no quantum beats, so we have to go to the dipole correlation to 
observe them. Using the quantum regression formula, we get
\begin{subequations}    \label{eq:EW-SpEm}
\begin{eqnarray} 	
\langle A_{13} (t_1) A_{31} (t_1+\tau) \rangle
	&=& \frac{1}{2} e^{-\gamma \tau/2} e^{-\gamma t_1},  \\
\langle A_{24} (t_1) A_{42} (t_1+\tau) \rangle 
	&=& \frac{1}{2} e^{-(\gamma/2 +i\delta) \tau} e^{-\gamma t_1} .
\end{eqnarray}
\end{subequations}
Inserting Eqs.~(\ref{eq:EW-SpEm}) into Eq.~(\ref{eq:spectra-pi}) 
gives us the spectrum Eq.~(\ref{eq:S_SE}). For numerical calculations, 
the populations and coherences are obtained from 
\begin{subequations}\label{eq:SEMatrixElements}
\begin{eqnarray} 	
\langle A_{11}(t) \rangle 
	&=& \left( [e^{\mathbf{M} t}]_{11} +[e^{\mathbf{M} t}]_{13} \right)/2 ,  \\
\langle A_{22}(t) \rangle 
&=& \left(  [e^{\mathbf{M} t}]_{31} +[e^{\mathbf{M} t}]_{33} \right)/2 ,  \\
\langle A_{13}(t) \rangle 
&=& \left( [e^{\mathbf{M} t}]_{21} +[e^{\mathbf{M} t}]_{23} \right)/2 =0 ,\\
\langle A_{24}(t) \rangle 
&=&  \left(  [e^{\mathbf{M} t}]_{41} +[e^{\mathbf{M} t}]_{43} \right)/2 =0 , 
\end{eqnarray}
\end{subequations}
where the subindices indicate the row and column of the matrix in brackets. 

To observe beats in resonance fluorescence, besides $\delta \neq 0$, 
we need both ground state populations to be nonzero initially, optimally 
\begin{eqnarray} 	\label{eq:initialCondBeats}
\langle \mathbf{Q}(0) \rangle_{\mathrm{RF}} 
	&=& \left( 0, 0, 0, 0, 0, 1/2,  0, 1/2 \right)^T .  
\end{eqnarray}
The populations and coherences are: 
\begin{subequations}    \label{eq:numPopsCohs}
\begin{eqnarray} 	
\langle A_{11}(t) \rangle 
&=& \left(  [e^{\mathbf{M} t}]_{16} +[e^{\mathbf{M} t}]_{18} \right)/2,  \\
\langle A_{22}(t) \rangle 
&=& \left(  [e^{\mathbf{M} t}]_{36} +[e^{\mathbf{M} t}]_{38} \right)/2,  \\
\langle A_{13}(t) \rangle 
&=& \left(  [e^{\mathbf{M} t}]_{26} +[e^{\mathbf{M} t}]_{28} \right)/2, \\
\langle A_{24}(t) \rangle 
&=& \left(   [e^{\mathbf{M} t}]_{46} +[e^{\mathbf{M} t}]_{48} \right)/2, 
\end{eqnarray}
\end{subequations}
giving the intensity, Eq.~(\ref{eq:time_Intensity_pi}), 
$2 I_{\pi}^{\mathrm{RF}} (t)  
	=  [e^{\mathbf{M} t}]_{16} + [e^{\mathbf{M} t}]_{18} 
	+ [e^{\mathbf{M} t}]_{36} + [e^{\mathbf{M} t}]_{38}$,  
displayed in Figs.~4 and 5 of Ref.~\cite{CSGA24}.

Following the procedure outlined above, the correlations are
\begin{subequations} 	\label{eq:CorrsInSpectra} 
\begin{eqnarray}
\langle A_{13}(t_1) A_{31}(t_1+\tau) \rangle 
&=& [e^{\mathbf{M} \tau}]_{55} \langle A_{11} (t_1) \rangle \nonumber \\
&&\quad\quad+[e^{\mathbf{M} \tau}]_{56} \langle A_{13} (t_1) \rangle ,\qquad \\
\langle A_{24}(t_1) A_{42}(t_1+\tau) \rangle 
&=& [e^{\mathbf{M} \tau}]_{77} \langle A_{22} (t_1) \rangle \nonumber \\
&&\quad\quad +[e^{\mathbf{M} \tau}]_{78} \langle A_{24} (t_1) \rangle ,\qquad 
\end{eqnarray}
\end{subequations}
where Eqs.~(\ref{eq:numPopsCohs}) (Eqs.~(\ref{eq:SEMatrixElements}) 
for spontaneous emission) are to be used, and then integrated as 
prescribed in Eq.~(\ref{eq:spectra-pi}) to find the spectrum
\begin{eqnarray}
\lefteqn{
S(\nu, t, \Gamma) = 2\Gamma \mathrm{Re}  \int_{0}^t dt_1  \, 
	e^{-\Gamma(t-t_1)} \int_{0}^{t-t_1} d\tau  \, e^{(\Gamma/2 -i \nu) \tau} } 
	\quad \nonumber \\
  &&\qquad  \times \left\{  [e^{\mathbf{M} \tau}]_{55} \langle A_{11} (t_1) \rangle 
	+[e^{\mathbf{M} \tau}]_{56} \langle A_{13} (t_1) \rangle  \right. \nonumber \\
	&&\qquad \left. +[e^{\mathbf{M} \tau}]_{77} \langle A_{22} (t_1) 
	+[e^{\mathbf{M} \tau}]_{78} \langle A_{24} (t_1) \rangle \right\}.\qquad\quad
\end{eqnarray}

We want to point to the temporal factorization of the functions of $t_1$ 
and $\tau$ in the correlations Eqs.~(\ref{eq:EW-SpEm}) and 
(\ref{eq:CorrsInSpectra}), permitting great simplification of the analytical 
and numerical evaluations of TDS. This observation allows to suggest the 
scalability to larger multilevel systems \cite{NOBC10} as long as the Bloch 
equations are homogeneous. Otherwise, the computations might be much 
more involved.


To ensure the reproducibility of our results for the reader's benefit, 
we share our source code in~\cite{HMCB+RRA}.


\section{Effect of Narrower and Broader Filter Bandwidth}
\label{App_Filter} 
\begin{figure}[h]
\includegraphics[scale=0.26]{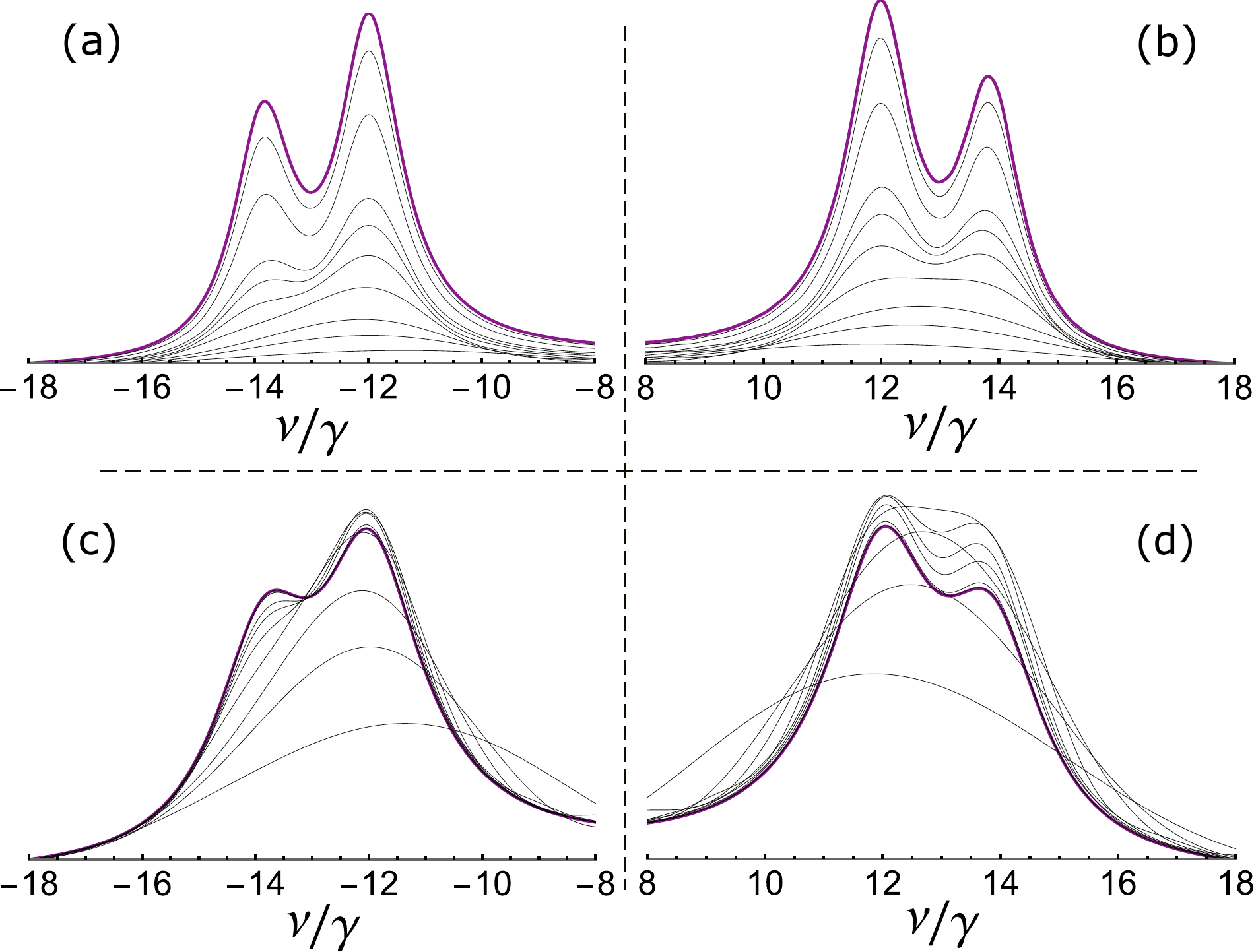} 
\caption{\label{Four_bands} 
Evolution of the left (left panel) and right (right panel) doublet sidebands 
of $S(\nu,t,\Gamma)$. Same parameters as Figs.~\ref{fig_full-left-ctr-G} 
and \ref{fig_right_sb} except for $\Gamma=0.1\gamma$ for (a) and (b), 
and $\Gamma=\gamma$ for (c) and (d). The purple (thicker) solid line is 
the trace at $\gamma t=20$, i.e., very close to the steady state.}
\end{figure}

Figure~\ref{Four_bands} shows the sidebands in the TDS 
of quantum beats in resonance fluorescence for finer 
($\Gamma=0.1 \gamma$, top panel) and coarser ($\Gamma=\gamma$, 
lower panel) filter resolution than those of Figs.~\ref{fig_full-left-ctr-G} and 
\ref{fig_right_sb}. The spectrum evolves more symmetrically with finer filter 
resolution, and the dip is better resolved. With a coarser filter, the 
asymmetry persists longer and the dip becomes shallow. 


\end{document}